\documentclass[reprint,aps,prl,floattix]{revtex4-1}

\usepackage{amsmath}
\usepackage{amssymb}
\usepackage{amsfonts}
\usepackage{graphicx}
\usepackage{float}
\usepackage{psfrag}

\begin{document}

\title{The role of geometric phase in the formation of electronic
  coherences at conical intersections}

\author{Simon P. Neville$^{1}$}
\author{Albert Stolow$^{1,2,3,4}$}
\author{Michael S. Schuurman$^{1,2}$}

\affiliation{$^1$National Research Council of Canada, 100 Sussex
  Drive, Ottawa, Ontario K1A 0R6, Canada}
\affiliation{$^2$Department of Chemistry and Biomolecular Sciences,
  University of Ottawa, 150 Louis Pasteur, Ottawa, Ontario, K1N 6N5,
  Canada}
\affiliation{$^3$Department of Physics, University of Ottawa, 150
  Louis Pasteur,Ottawa, ON K1N 6N5 Canada}
\affiliation{$^4$University of Ottawa - National Research Council
  Joint Centre for Extreme Photonics, Ottawa ON K1A 0R6, Canada}

\begin{abstract}
  The direct observation of non-adiabatic dynamics at conical
  intersections is a long-standing goal of molecular physics. Novel
  time-resolved spectroscopies have been proposed which are sensitive
  to electronic coherences induced by the passage of an excited state
  wavepacket through a region of conical intersection.  Here we
  demonstrate that inclusion of the geometric phase effect, and its
  manifestations, is essential for the correct description of the
  transient electronic coherences that may or may not develop. For
  electronic states of different symmetry, electronic coherences are
  suppressed by the geometric phase. Conversely, for states of the
  same symmetry, appreciable electronic coherences are possible, but
  their magnitude depends on both the topography of and direction of
  approach to the conical intersection. These general results have
  consequences for all studies of electronic coherences at conical
  intersections.
\end{abstract}

\maketitle

The concept of seams of conical intersections (CIs) is central to our
understanding of ultrafast electronic relaxation from molecular
excited states. The points forming the seam correspond to degeneracies
between adiabatic electronic states, providing the pathways by which
internal conversion may occur on vibrational
timescales\cite{schuurman_conical_intersections_review}. The existence
of such ultrafast non-adiabatic pathways is of immense importance in
many areas of photoinduced molecular dynamics, including vision,
photosynthesis, the photostability of biomolecules, and light
harvesting systems. There is thus global interest in developing
experimental methods to directly observe populations and coherences
induced by CI mediated molecular dynamics.

Ultrafast spectroscopic probes of dynamics at CIs can be broadly split
into two categories. The first encompasses methods which are sensitive
to electronic state \textit{population} dynamics of a vibronic
wavepacket as it transits the region of strong non-adiabatic coupling
surrounding a CI seam. These include, but are not limited to,
ultrafast pump-probe spectroscopies such as transient absorption or
time-resolved photoelectron
spectroscopy\cite{ethylene_faraday_ours,ethylene_trxas_prl_ours,leone_trxas_acc_chem_res,schuurman_conical_intersections_review,underwood_trpes_adv_chem_phys,stolow_trpes_review_2004}. The
second category is comprised of (generally non-linear) spectroscopic
methods which are sensitive to the transient electronic
\textit{coherences} induced as a wavepacket passes through a region
encompassing a
CI\cite{mukamel_conical_intersections_x-ray_raman,mukamel_chem_rev_2017,mukamel_thiophenol_truecars,makhija_anisotropies_nh3,domcke_ultrafast_spectroscopy_cis}.

For a coherence to exist between two electronic states, two conditions
must be satisfied: (i) both electronic states must be appreciably
populated; (ii) the nuclear components of the wavepacket on each
adiabatic state must have appreciable overlap. Since the nonadiabatic
coupling becomes large in the region around a CI, population is
readily transferred between the electronic states, ensuring the first
criterion is met. However, we show here that the geometric
phase\cite{izmaylov_geometric_phase_review} effect can lead to zero,
or near-zero, overlaps of the nuclear components of the wavepacket on
different electronic states. For example, the geometric phase effect
is found to suppress the formation of coherences between electronic
states of different symmetry. Conversely, for CIs between electronic
states of the same symmetry, large coherences may form. However, the
magnitude of the electronic coherences thus formed strongly depends on
both the CI topography and the direction of approach of the
wavepacket.

We begin with a brief exposition of the problem using a minimal model
which captures all the fundamental physical effects involved. Consider
the excitation of a molecule from its ground electronic state to an
excited state manifold spanned by two adiabatic states $|
\psi_{1}^{(a)} \rangle$ and $| \psi_{2}^{(a)} \rangle$. We consider
the case where only $| \psi_{2}^{(a)} \rangle$ is initially
populated. The non-stationary vibronic wavepacket produced in $|
\psi_{2}^{(a)} \rangle$ may evolve towards a CI between the two
states, leading to internal conversion and population transfer to $|
\psi_{1}^{(a)} \rangle$. This process is more easily modeled by
switching to a set of diabatic states, $| \psi_{1}^{(d)} \rangle$ and
$| \psi_{2}^{(d)} \rangle$, related to the adiabatic states by a
unitary transformation,

\begin{equation}\label{eq:adt}
  \begin{bmatrix}
    | \psi_{1}^{(a)} \rangle \\
    | \psi_{2}^{(a)} \rangle
  \end{bmatrix}
  =
  \begin{bmatrix}
    \cos{\theta} & \sin{\theta} \\
    -\sin{\theta} & \cos{\theta} \\
  \end{bmatrix}
  \begin{bmatrix}
    | \psi_{1}^{(d)} \rangle \\
    | \psi_{2}^{(d)} \rangle
  \end{bmatrix},
\end{equation}

\noindent
where $\theta$ is the (geometry-dependent) adiabatic-to-diabatic
transformation (ADT) angle.

In the diabatic representation, the Hamiltonian matrix reads

\begin{equation}\label{eq:diab_ham}
  \boldsymbol{H} = \hat{T}_{n} \boldsymbol{1}_{2}
  + \begin{bmatrix}
    W_{11} & W_{21} \\
    W_{12} & W_{22}
    \end{bmatrix},
\end{equation}

\noindent
where $\hat{T}_{n}$ is the nuclear kinetic energy operator and
$W_{ij}$ are the elements of the diabatic potential matrix, the
off-diagonal elements of which account for the non-adiabatic coupling
between the two electronic states.

First, consider a system comprised of two electronic states of
\textit{different} symmetry. The simplest possible model correctly
accounting for the underlying physics and symmetries of this problem
is a two-mode linear vibronic coupling (LVC) Hamiltonian,
$\hat{H}_{LVC}$, expressed in terms of a totally symmetric tuning mode
$q_{t}$ and a coupling mode $q_{c}$:

\begin{equation}\label{eq:lvc_2mode}
  \begin{aligned}
  \boldsymbol{H}_{LVC} &= \left[ \hat{T}_{n} + \frac{1}{2}
    \left(\omega_{t} q_{t}^{2} + \omega_{c} q_{c}^{2}\right) \right]
  \boldsymbol{1}_{2} \\
  &+ \begin{bmatrix}
    \kappa_{1} q_{t} & \lambda q_{c} \\
    \lambda q_{c} & \Delta + \kappa_{2} q_{t}
    \end{bmatrix}.
  \end{aligned}
\end{equation}

\noindent
This model describes a system of shifted, coupled harmonic oscillators
with frequencies $\omega_{t/c}$ and a CI located at
$\boldsymbol{Q}_{CI} = (q_{t}=\Delta/(\kappa_{1}-\kappa_{2}),
q_{c}=0)$. Note that the coupling mode in this case is non-totally
symmetric, with a symmetry given by the direct product of the
irreducible representations generated by the two electronic states.

The quantity of interest here is the magnitude of the coherence
between the adiabatic electronic states as the excited state
wavepacket passes through the CI. Let $\Psi(t)$ denote the total
vibronic wavepacket, which can be expressed in the Born-Huang
framework as

\begin{equation}
  \Psi(t) = \sum_{j=1}^{2} | \psi_{j}^{(a)} \rangle
  \chi_{j}^{(a)}(q_{t},q_{c},t).
\end{equation}

\noindent
where the $\chi_{j}^{(a)}$ are the adiabatic nuclear wavepackets.

The magnitude of the coherence between the electronic states is given
by the absolute value of off-diagonal element of the reduced
electronic density matrix in the adiabatic representation,

\begin{equation}
  \rho_{12}^{(a)}(t) = \left\langle \Psi(t) \middle| \psi_{1}^{(a)}
  \right\rangle \left\langle \psi_{2}^{(a)} \middle| \Psi(t)
  \right\rangle = \left\langle \chi_{1}^{(a)}(t) \middle|
  \chi_{2}^{(a)}(t) \right\rangle,
\end{equation}

As stated above, in order for $|\rho_{12}^{(a)}(t)|$ to be large in
value, two conditions must be met: both electronic states must be
populated and the nuclear wavepackets, $\chi_{j}^{(a)}$, on each
adiabatic state must have appreciable overlap. The strong
non-adiabatic coupling in the CI region ensures that the first
condition is met. The second condition, however, is generally
\textit{not} satisfied when the electronic states are of different
symmetry.

The reason for poor overlap between the two adiabatic nuclear
wavepackets of concern is best seen by re-writing $\rho_{12}^{(a)}(t)$
as

\begin{equation}\label{eq:rho12}
  \rho_{12}^{(a)}(t) = A_{12}(t) + B_{12}(t),
\end{equation}

\begin{equation}\label{eq:A12}
  2 A_{12}(t) = \left\langle \chi_{1}^{(d)} \middle| \sin 2 \theta
  \middle| \chi_{1}^{(d)} \right\rangle + \left\langle \chi_{2}^{(d)}
  \middle| \sin 2\theta \middle| \chi_{2}^{(d)} \right\rangle
\end{equation}

\begin{equation}\label{eq:B12}
    2 B_{12}(t) = \left\{ \left\langle \chi_{1}^{(d)} \middle|
    \chi_{2}^{(d)} \right\rangle + \left\langle \chi_{1}^{(d)}
    \middle| \cos 2\theta \middle| \chi_{2}^{(d)} \right\rangle
    \right\} + h.c.
\end{equation}

\noindent
where $\theta$ is the ADT angle, and the $\chi_{j}^{(d)}$ are the
\textit{diabatic} nuclear wavepackets. In the following, we will refer
to $A_{12}$ as the ``on-diagonal'' contribution to $\rho_{12}^{(a)}$,
and $B_{12}$ as the ``off-diagonal'' contribution.

We can show that both the $A_{12}$ and $B_{12}$ terms in Equation
\ref{eq:A12} vanish using the following simple symmetry arguments:

\begin{enumerate}
  
\item Because the two electronic states are of different symmetry, the
  coupling mode $q_{c}$ will be non-totally symmetric. This implies
  that $W_{11}$ and $W_{22}$ are even functions with respect to
  $q_{c}$, while $W_{12}$ is an odd function of $q_{c}$.

\item Given the even symmetry of $W_{22}$ with respect to $q_{c}$, and
  assuming vertical excitation from the ground state, $\chi_{2}^{(d)}$
  will be an even function of $q_{c}$.

\item Given the odd symmetry of $W_{12}$ with respect to $q_{c}$ and
  the even symmetry of $\chi_{2}^{(d)}$, the nuclear wavepacket
  $\chi_{1}^{(d)}$ formed on the lower electronic state will have a
  node along the coupling mode $q_{c}$, and thus will be an odd
  function of this coordinate.

\item Using the relation

  \begin{equation}
    2 \theta = \arctan \frac{2W_{12}}{W_{22}-W_{11}},
  \end{equation}

  it can be shown that the functions $\sin{2\theta(q_{t},q_{c})}$ and
  $\cos{2\theta(q_{t},q_{c})}$ are odd and even with respect to
  $q_{c}$, respectively (see the Supplementary Information for more
  details).
  
\end{enumerate}

Given these symmetries, we see that the integrands in both Equations
\ref{eq:A12} and \ref{eq:B12} are all overall odd with respect to
$q_{c}$ and vanish in this two-mode LVC model. We conclude that
electronic coherences are suppressed for CIs between states of
different symmetry.

As detailed in the Supplementary Information, the symmetries of
$\sin{2\theta(q_{t},q_{c})}$ and $\cos{2\theta(q_{t},q_{c})}$ are
intimately linked to the geometric phase effect. Further, the
geometric phase effect will exist if the leading term in the diabatic
coupling is odd with respect to $q_{c}$. This symmetry is realized in
the LVC model above and is responsible for the node in
$\chi_{1}^{(d)}$. This is the reason why the geometric phase effect is
present in the case of conical intersections, but is absent for
glancing (e.g. Renner-Teller) intersections. Thus, the suppression of
coherences between electronic states of different symmetry is best
understood as simply a consequence of the geometric phase effect.

To demonstrate this, we note that by forcing the diabatic coupling to
contain only even terms with respect to the coupling mode(s), the
geometric phase effect may be ``turned off''
\cite{izmaylov_geometric_phase_diabatic_definition}. This can be
achieved by replacing $W_{12}$ in Equation \ref{eq:diab_ham} by its
absolute value: $W_{12} \rightarrow |W_{12}|$. In the following, we
denote the physically correct Hamiltonian as $\hat{H}_{\text{wGP}}$
(i.e. with geometric phase), and the Hamiltonian with $W_{12}$
replaced by its absolute value by $\hat{H}_{\text{noGP}}$ (i.e. no
geometric phase). Importantly, using $\hat{H}_{\text{noGP}}$ has no
effect on the adiabatic potential energy surfaces. However, it allows
us to clearly demonstrate the role of the geometric phase effect on
the formation of electronic coherences. In the following, we perform
wavepacket propagations using both $\hat{H}_{\text{wGP}}$ and
$\hat{H}_{\text{noGP}}$ and track the resulting changes in the values
of $|\rho_{12}^{(a)}(t)|$.

As a representative system, we choose the excited state dynamics of
pyrazine, using a four-mode model to describe the coupled dynamics of
the $B_{3u}(n \pi^{*})$ and $B_{2u}(\pi \pi^{*})$ states (adapted from
Reference \citenum{domcke_pyrazine_path_integral}). This well-known
model correctly reproduces the short-time excited state dynamics of
pyrazine, including the passage of the excited state wavepacket
through the CI between the $B_{3u}(n\pi^{*})$ and $B_{2u}(\pi\pi^{*})$
states. We begin by considering the symmetries of the nuclear
wavepackets $\chi_{1}^{(d)}$ and $\chi_{2}^{(d)}$ formed following
vertical excitation to the bright $B_{3u}(n\pi^{*})$ state. These
symmetries are most readily discerned from the phase angles
$\zeta_{j}$ defined {\it via} the the polar representation of the
diabatic nuclear wavepackets,

\begin{equation}
  \chi_{j}^{(d)}(\boldsymbol{Q},t) = r_{j}(\boldsymbol{Q},t) e^{i
    \zeta_{j}(\boldsymbol{Q},t)}.
\end{equation}

\begin{figure*}
  \begin{minipage}{0.5\textwidth}
    \centering
    \includegraphics[width=9cm]{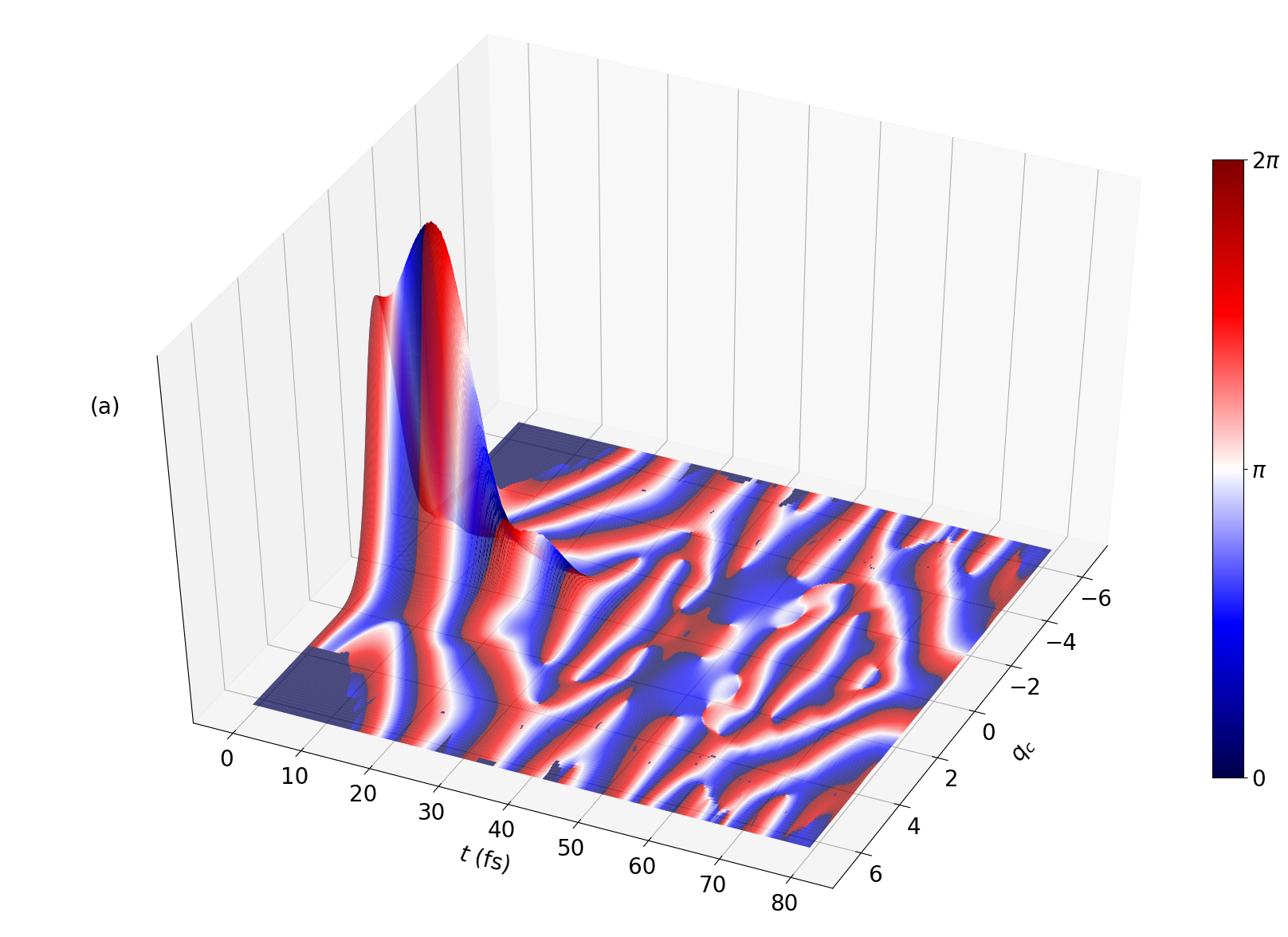}
  \end{minipage}\hfill
  \begin{minipage}{0.5\textwidth}
    \centering
    \includegraphics[width=9cm]{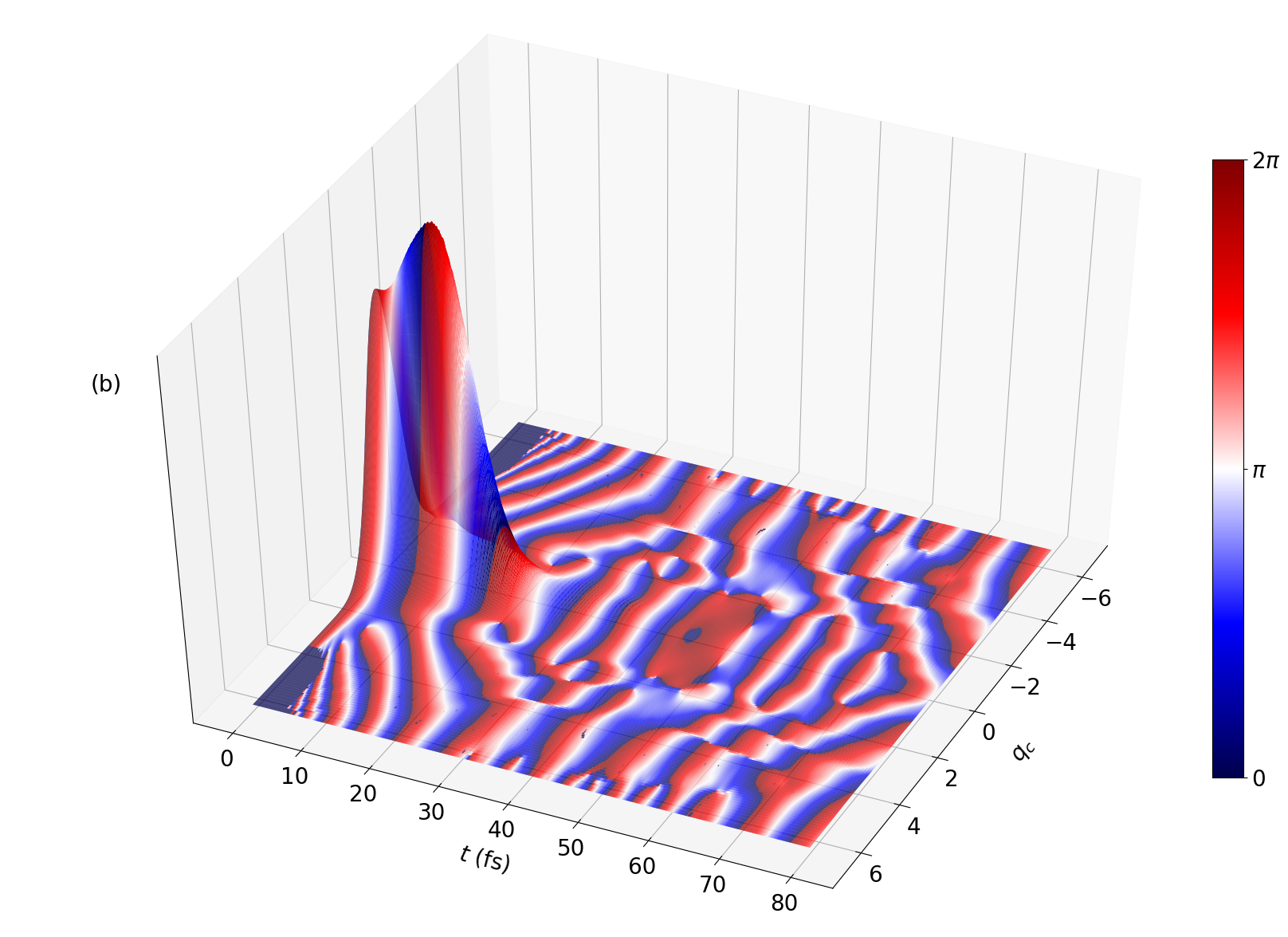}
  \end{minipage}
  \begin{minipage}{0.5\textwidth}
    \centering
    \includegraphics[width=10cm]{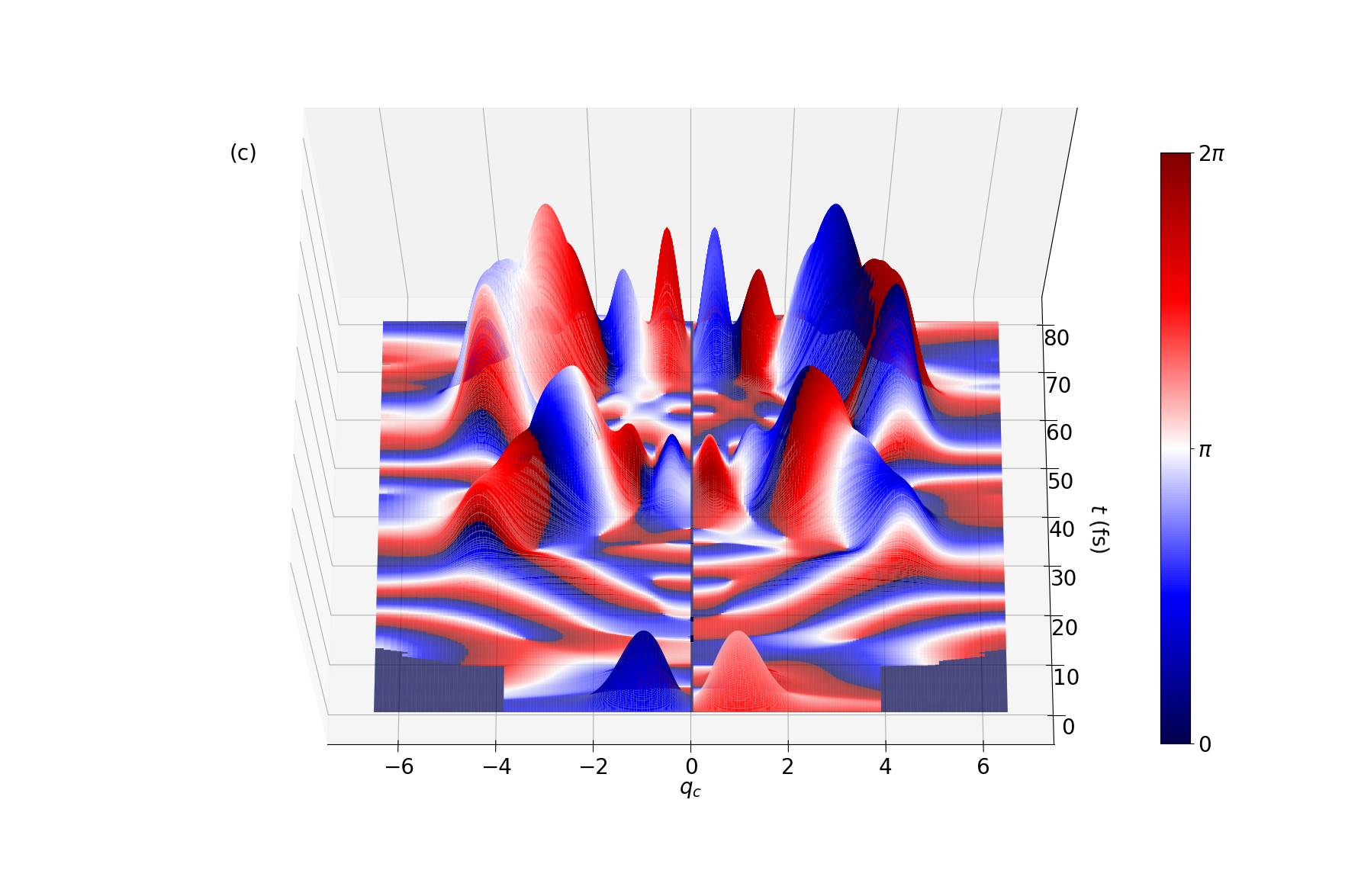}
  \end{minipage}\hfill
  \begin{minipage}{0.5\textwidth}
    \centering
    \includegraphics[width=10cm]{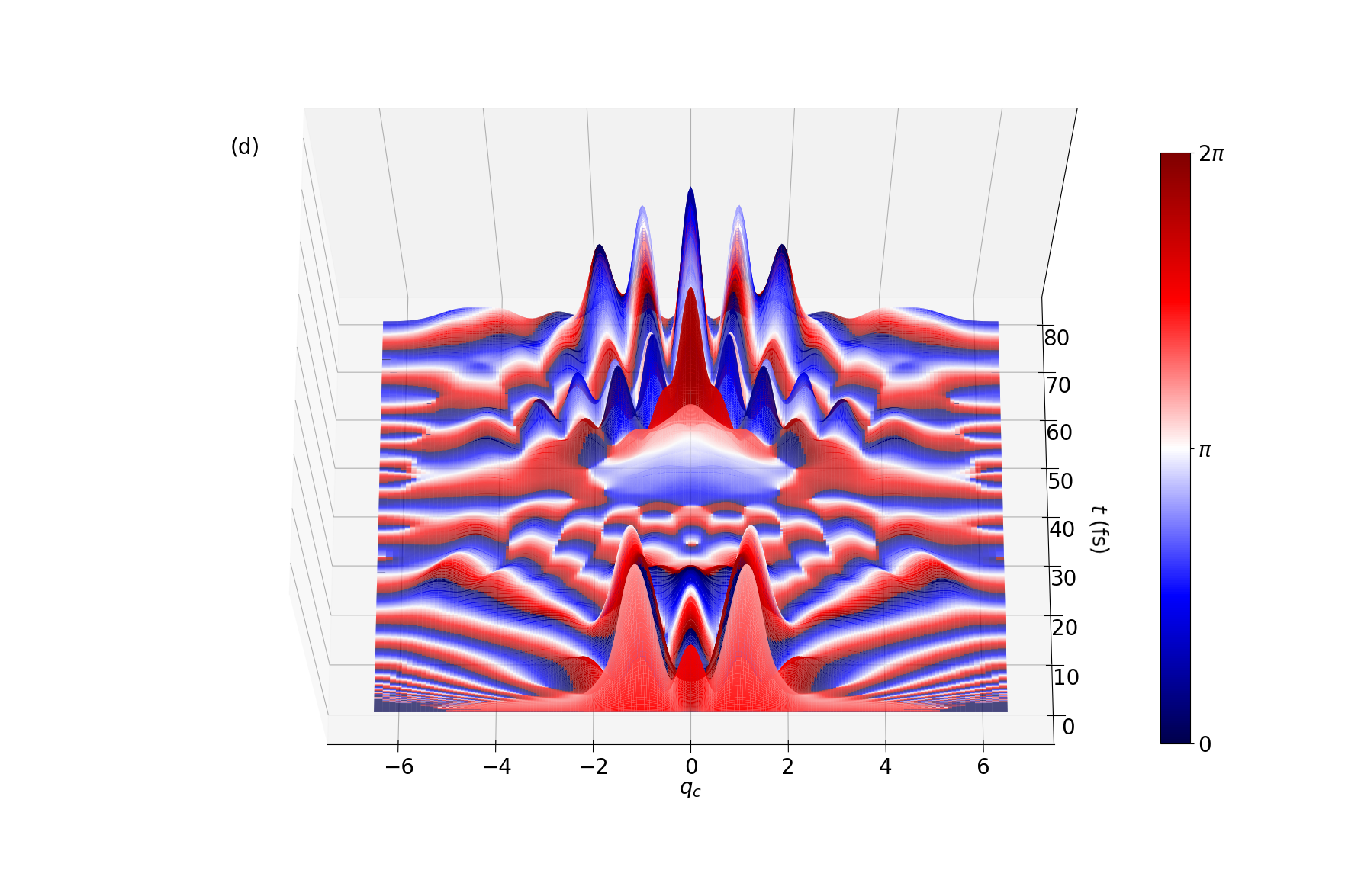}
  \end{minipage}
  \caption{Dynamics at a conical intersection in pyrazine, showing the
    nuclear wavepacket evolution in the coupled $B_{3u}(n \pi^{*})$
    and $B_{2u}(\pi \pi^{*})$ states. The the magnitude in the
    \textbf{z} direction in each plot shows the squared absolute
    values $|\chi_{j}^{(d)}|^{2}$ of the diabatic nuclear
    wavepackets. The phase angle $\zeta_{j}$ of the wavepackets is
    encoded in the color maps. (a) Upper diabatic state using
    $\hat{H}_{\text{wGP}}$. (b) Upper diabatic state using
    $\hat{H}_{\text{noGP}}$. (c) Lower diabatic state using
    $\hat{H}_{\text{wGP}}$. (d) Lower diabatic state using
    $\hat{H}_{\text{noGP}}$. In all cases, the phase angle is plotted
    along the coupling mode $q_{c}$ with all other nuclear degrees set
    to their time-dependent centroid values.}
    \label{fig:phase_angle}
\end{figure*}

\noindent
A phase angle satisfying $\zeta_{j}(q_{c}) = \zeta_{j}(-q_{c})$
corresponds to an even function of $q_{c}$, while $\zeta_{j}(q_{c}) =
\zeta_{j}(-q_{c}) \pm \pi$ implies and odd function of $q_{c}$. Shown
in Figure \ref{fig:phase_angle} are the squared absolute values
$|\chi_{j}^{(d)}|^{2}$ of the diabatic nuclear wavepackets colored by
the phase angles $\zeta_{j}$. In these plots, $|\chi_{j}^{(d)}|^{2}$
and $\zeta_{j}$ are shown plotted along $q_{c}$ with all other modes
set to their time-evolving centroid values.

As shown by the amplitude (i.e. the magnitude in the \textbf{z}
direction) of the nuclear wavepackets in Figures
\ref{fig:phase_angle}a and \ref{fig:phase_angle}b, it is the upper
diabatic state which is primarily populated at early times. However,
this population is quickly depleted following the initial passage
through the CI region at around 40 fs. Conversely, the lower diabatic
state rapidly accumulates population at this time, as evinced by
Figures \ref{fig:phase_angle}c and \ref{fig:phase_angle}d. These
observations regarding the state populations apply equally well to
both the $\hat{H}_{\text{wGP}}$ and $\hat{H}_{\text{noGP}}$
simulations. That is, the geometric phase is of little consequence for
the population dynamics, as shown below in Figure \ref{fig:pyrazine}a
(\textit{vide infra}).

In contrast, there are significant differences in the \textit{phase}
of nuclear wave packets determined from $\hat{H}_{\text{wGP}}$ and
$\hat{H}_{\text{noGP}}$.  The phase angle for $\chi_{2}^{(d)}$ is an
even function of $q_{c}$ for both models, as can be seen in Figures
\ref{fig:phase_angle}a and \ref{fig:phase_angle}b. As discussed above,
the use of the physically correct Hamiltonian $\hat{H}_{\text{wGP}}$
causes a node to form in $\chi_{1}^{(d)}$, resulting in it being an
odd function of $q_{c}$ (see Figure \ref{fig:phase_angle}c). Using the
unphysical Hamiltonian $\hat{H}_{\text{noGP}}$ results in
$\chi_{1}^{(d)}$ being an even function of $q_{c}$, as shown in Figure
\ref{fig:phase_angle}d. Thus, when coupled with the above discussed
symmetry properties of the ADT angle (see Supplementary Information),
the use of $\hat{H}_{\text{wGP}}$ results in the suppression of the
electronic coherences in this system, whilst the use of
$\hat{H}_{\text{noGP}}$ results in (spurious) electronic coherences of
large magnitude. This is clearly seen in Figure \ref{fig:pyrazine}b,
where we show the magnitudes $|\rho_{12}^{(a)}(t)|$ of the electronic
coherences formed when using both $\hat{H}_{\text{wGP}}$ and
$\hat{H}_{\text{noGP}}$, where the latter are completely suppressed.
Significantly, although the magnitude of the electronic coherences
formed using $\hat{H}_{\text{wGP}}$ and $\hat{H}_{\text{noGP}}$ are
entirely different, the population dynamics are remarkably similar, as
illustrated in Figure \ref{fig:pyrazine}a. This result shows that
although neglecting the geometric phase effect may have only mild
consequences for the simulation of electronic state population
dynamics, it must be properly accounted for when studying electronic
coherences.


We now consider the case of CIs between two electronic states of the
same symmetry. Here, the coupling mode $q_{c}$ will generate the
totally symmetric irreducible representation of the point group of the
molecule. Hence, there can exist non-zero gradients of diabatic
potentials with respect to $q_{c}$, and the diabatic nuclear
wavepackets $\chi_{1/2}^{(d)}$ will no longer necessarily be even or
odd functions of it, Accordingly, neither the on- or off-diagonal
contributions to $\rho_{12}^{(a)}$ necessarily vanish, and it is
possible for appreciable electronic coherences to form.

The magnitude of the electronic coherence formed in this same symmetry
case depends on both the topography of the CI and the direction of
approach of the wavepacket to the CI. To see this, consider the
first-order expansion of the diabatic potentials about the CI point,
$\boldsymbol{X}_{CI}$, in terms of intersection-adapted coordinates
$x$ and $y$\cite{atchity_intersection_adapted_coords},

\begin{equation}\label{eq:W1}
  \boldsymbol{W}^{(1)}(x,y) = \left( s_{x} x + s_{y} y \right)
  \boldsymbol{1}_{2} +
  \begin{bmatrix}
    -g x & h y \\
    h y & g x
  \end{bmatrix},
\end{equation}

\noindent
where $g$ and $h$ are the norms of the gradient difference and
non-adiabatic coupling vectors evaluated at the CI point,
respectively. The terms $s_{x/y}$ are the gradients of the average
energy with respect to $x$ and $y$ at the CI point, and determine
whether the CI is ``sloped'' or
``peaked''\cite{atchity_intersection_adapted_coords,yarkony_conical_intersection_topography}. In
Equation \ref{eq:W1}, the diabatic and adiabatic representations are
equal at the CI point $\boldsymbol{X}_{CI}$. In order to apply
symmetry arguments analogous to the different state symmetry case
above, we require that the linear component of the diabatic coupling
introduces a node at the centre of $\chi_{1}^{(d)}$. This is achieved
by transforming to a different, but entirely equivalent, diabatic
representation in which the diabatic and adiabatic representations are
equal at the centre of the initial wavepacket $\Psi(t=0)$. Note that
this is possible because the ADT is only defined up to a constant
unitary transformation. Let $\theta_{0}$ denote the ADT angle of the
original diabatic representation evaluated at the centre of the
initial wavepacket. Then, as detailed in the Supplementary
Information, the first-order potential in the new diabatic
representation takes the form

\begin{widetext}
  \begin{equation}\label{eq:W1_trans}
    \boldsymbol{W}^{(1)}(x,y) = \left( s_{x} x + s_{y} y \right)
    \boldsymbol{1}_{2}
    + \begin{bmatrix}
      -\cos{\left( 2\theta_{0} \right)} g x + \sin{\left( 2\theta_{0}
        \right)} h y & \cos{\left( 2\theta_{0} \right)} h y +
      \sin{\left( 2\theta_{0} \right)} g x \\
      \cos{\left( 2\theta_{0} \right)} h y + \sin{\left( 2\theta_{0}
        \right)} g x & \cos{\left( 2\theta_{0} \right)} g x -
      \sin{\left( 2\theta_{0} \right)} h y
    \end{bmatrix}.
  \end{equation}
\end{widetext}

To form an electronic coherence in this first-order vibronic coupling
model, there must be both on- and off-diagonal elements which are
non-vanishing with respect to either $x$ or $y$. The diabatic nuclear
wavepackets $\chi_{1/2}^{(d)}$ will then be neither even nor odd with
respect to both nuclear degrees of freedom, allowing for non-zero
values of both the on- and off-diagonal contributions to
$\rho_{12}^{(a)}$. Importantly, from Equation \ref{eq:W1_trans} we see
that this criterion will be satisfied if either: (i) $s_{y} \ne 0$,
and/or; (ii) $ \sin{\left( 2\theta_{0} \right)} \ne 0$. The parameter
$s_{y}$ determines the tilt of the CI axis along the non-adiabatic
coupling direction\cite{atchity_intersection_adapted_coords}, whereas
$\theta_{0}$ is determined by the position of the initial wavepacket
relative to the CI point. In a trajectory-based treatment of the
nuclear dynamics, this initial position would be analogous to the
``direction of approach'' to the CI. In the limiting case of a peaked
(i.e., non-tilted) CI and an initial wavepacket displaced from the CI
purely along the gradient difference direction, the dominant
first-order contributions to the electronic coherence will, again,
vanish.

\begin{figure*}
  \begin{center}
    \includegraphics[width=14.0cm,angle=0]{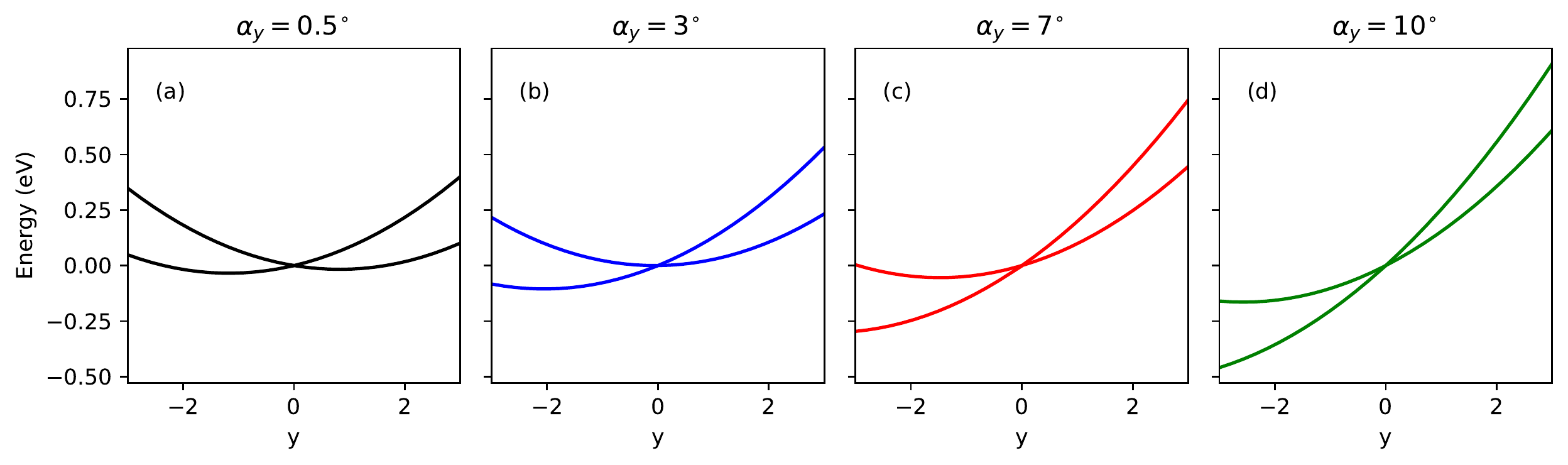}
    \includegraphics[width=7.0cm,angle=0]{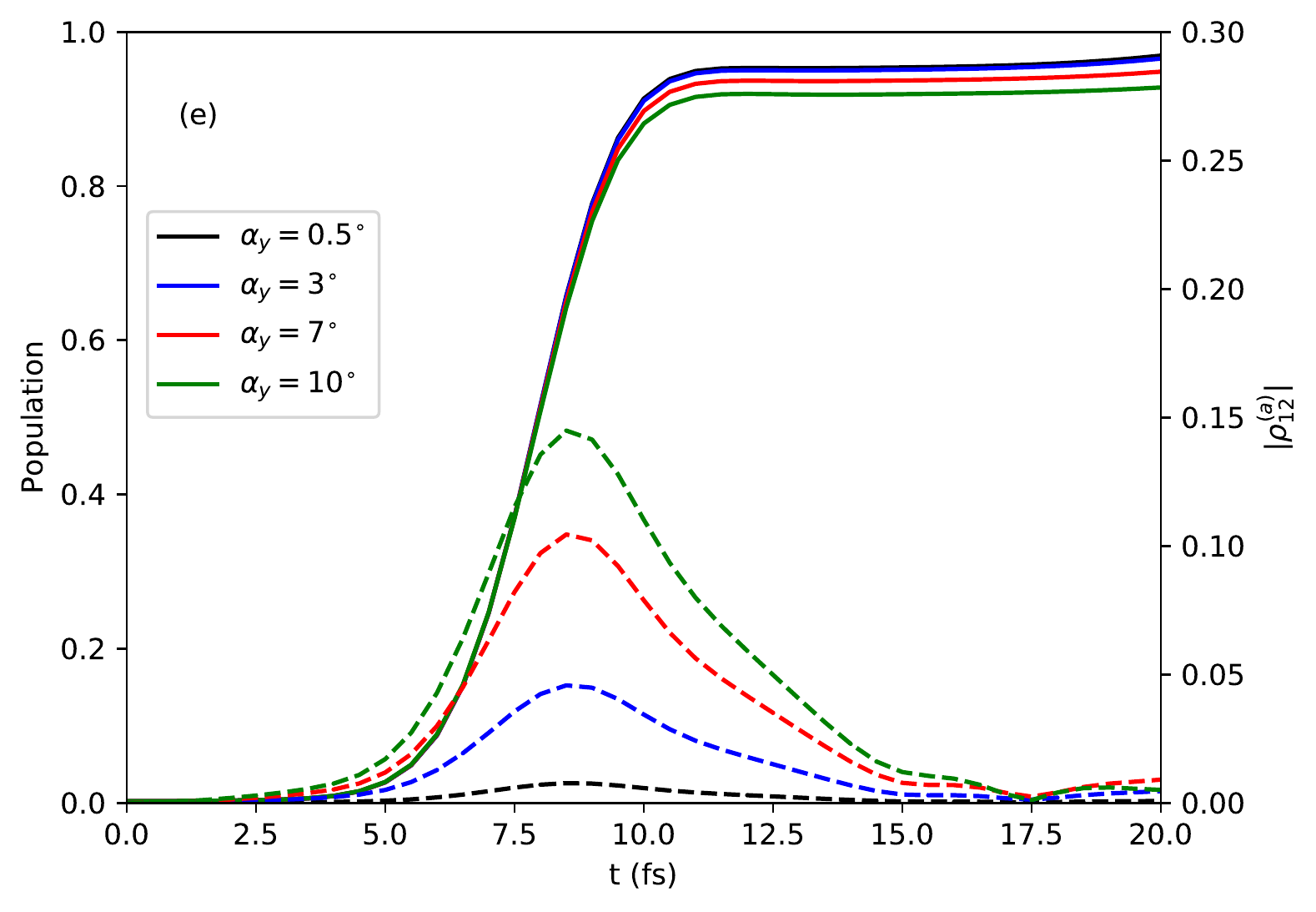}
    \includegraphics[width=7.0cm,angle=0]{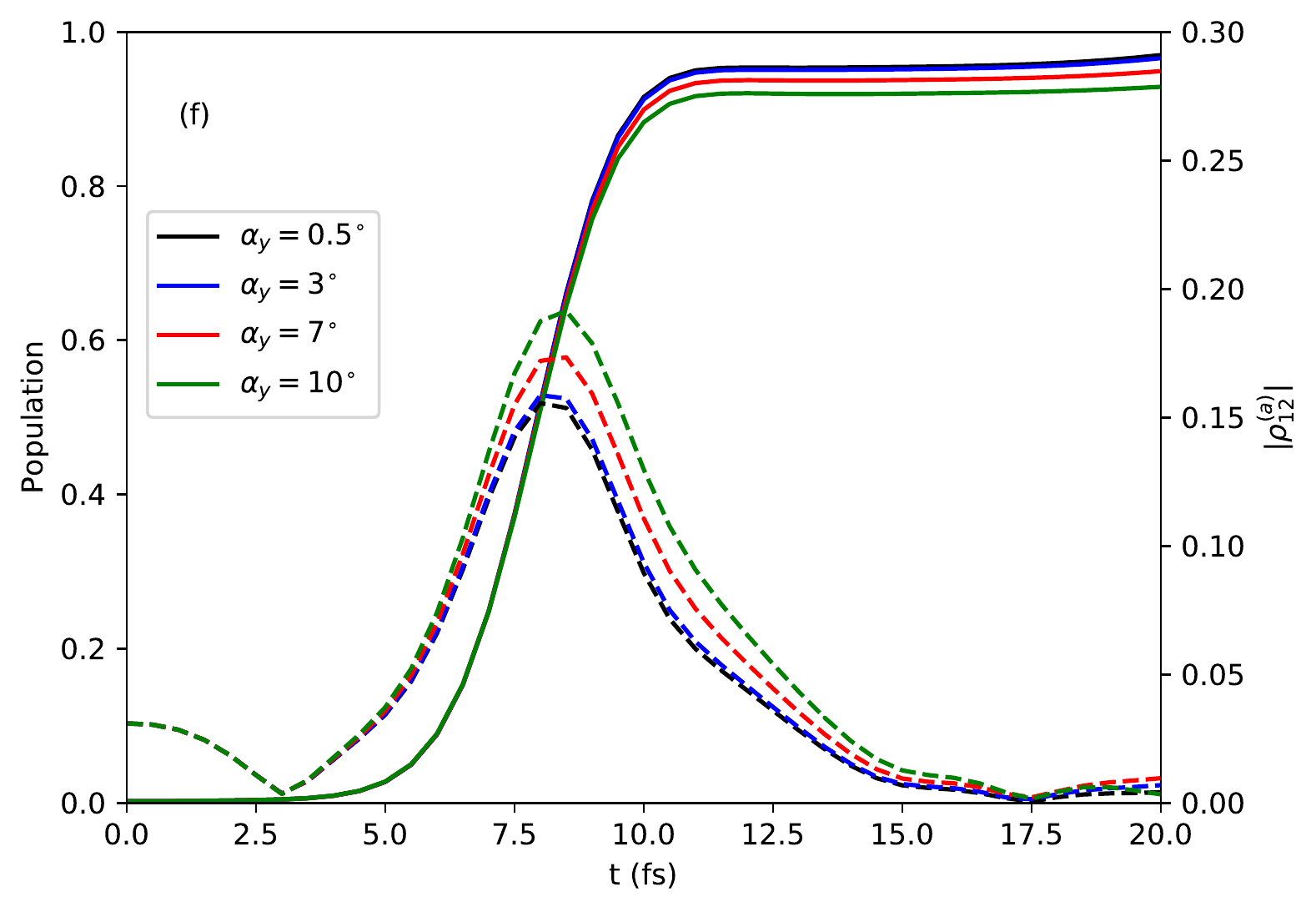}
    \caption{Conical intersections, population and electronic
      coherence dynamics in a two-mode, two-state model system as a
      function of the CI tilt angle $\alpha_{y}$ along the
      non-adiabatic coupling direction $y$. In panels (a) to (d) we
      show adiabatic potential surfaces as a function of increasing
      tilt angle $\alpha_{y}$. (e) Populations (solid lines) and
      electronic coherences (dashed lines) computed calculated using
      the physically correct $\hat{H}_{\text{wGP}}$. (f) Populations
      and coherences computed using the physically incorrect
      $\hat{H}_{\text{noGP}}$.  It can be seen that the omission of
      geometric phase leads to artificially larger electronic
      coherences for all tilt angles. In both cases, the population
      and coherence dynamics correspond to an initial wavepacket
      displaced from the CI point only along the gradient difference
      direction, thus isolating the effects due to CI topography
      alone.}
    \label{fig:ci_alphay}
  \end{center}
\end{figure*}

To illustrate the effects of CI topography on electronic coherences,
we consider the case where the centre of the initial wavepacket is
displaced from the CI point purely along the gradient difference
direction. In this case, the magnitude of the electronic coherence
will be determined by the tilt of the CI axis along the non-adiabatic
coupling direction $y$, and thus, by the parameter $s_{y}$ in Equation
\ref{eq:W1_trans}. A two-mode, two-state LVC Hamiltonian was
constructed to describe ultrafast, gradient-directed internal
conversion through a CI (see the Supplementary Information for the
parameters used). The parameter $s_{y}$ was varied to yield tilt
angles, $\alpha_{y}$, of $0.5^{\circ}$, $3^{\circ}$, $7^{\circ}$ and
$10^{\circ}$ along the $y$ direction. The conical intersections for
these tilt angles are shown in Figures \ref{fig:ci_alphay}a through
\ref{fig:ci_alphay}d. For $\alpha_{y}=0.5^{\circ}$, the CI is almost
peaked along $y$. Upon increasing $\alpha_{y}$, the slope of the CI
along $y$ gradually increases.
As illustrated in Figure \ref{fig:ci_alphay}d, the magnitude of the
electronic coherence is negligible for a nearly peaked CI
($\alpha_{y}=0.5^{\circ}$), and grows with increasing tilt angle along
$y$. Again, the adiabatic state population dynamics are found to be
only very weakly affected by the tilt of the CI along $y$. Finally, we
show in Figure \ref{fig:ci_alphay}e the same electronic state
populations and coherences, but with the geometric phase effect
removed from the model Hamiltonian (\textit{via} the use of
$\hat{H}_{\text{noGP}}$). Strikingly, the removal of the geometric
phase effect results - erroneously - in large magnitude electronic
coherences for all tilt angles. This again serves to highlight the
importance of correctly accounting for the geometric phase in any
simulation of electronic coherences at CIs.

In summary, we have explored the factors affecting the formation of
electronic coherences as a wavepacket passes through a
CI. Specifically, we have shown that the explicit consideration of the
geometric phase effect is essential for a qualitatively correct
description of the coherences which may or may not form in the
vicinity of a CI.  In the case of two electronic states of different
symmetry, geometric phase is responsible for the suppression of
electronic coherences around a CI. For the unavoidable case of a CI
between electronic states of the same symmetry, electronic coherences
may in general form. However, their magnitude depends on both the
topography of the CI and the direction of approach of the wavepacket
to it. These results will help to identify molecular systems for the
experimental study of electronic coherences in dynamics at CIs, an
opportunity identified in a recent road-map on ultrafast X-ray
science\cite{doe_report}. We emphasize that a proper accounting of the
geometric phase effect is required in any theoretical study of
electronic coherences induced by nuclear motion near CIs. This
recognition should result in the simulations required to guide
experimental efforts to identify unique signatures of CI dynamics.

\section{Acknowledgments}
The authors thank A. F. Izmaylov, I. G. Ryabikin, and
L. Joubert-Doriol for helpful discussions.

\begin{figure*}
  \begin{center}
    \includegraphics[width=7.0cm,angle=0]{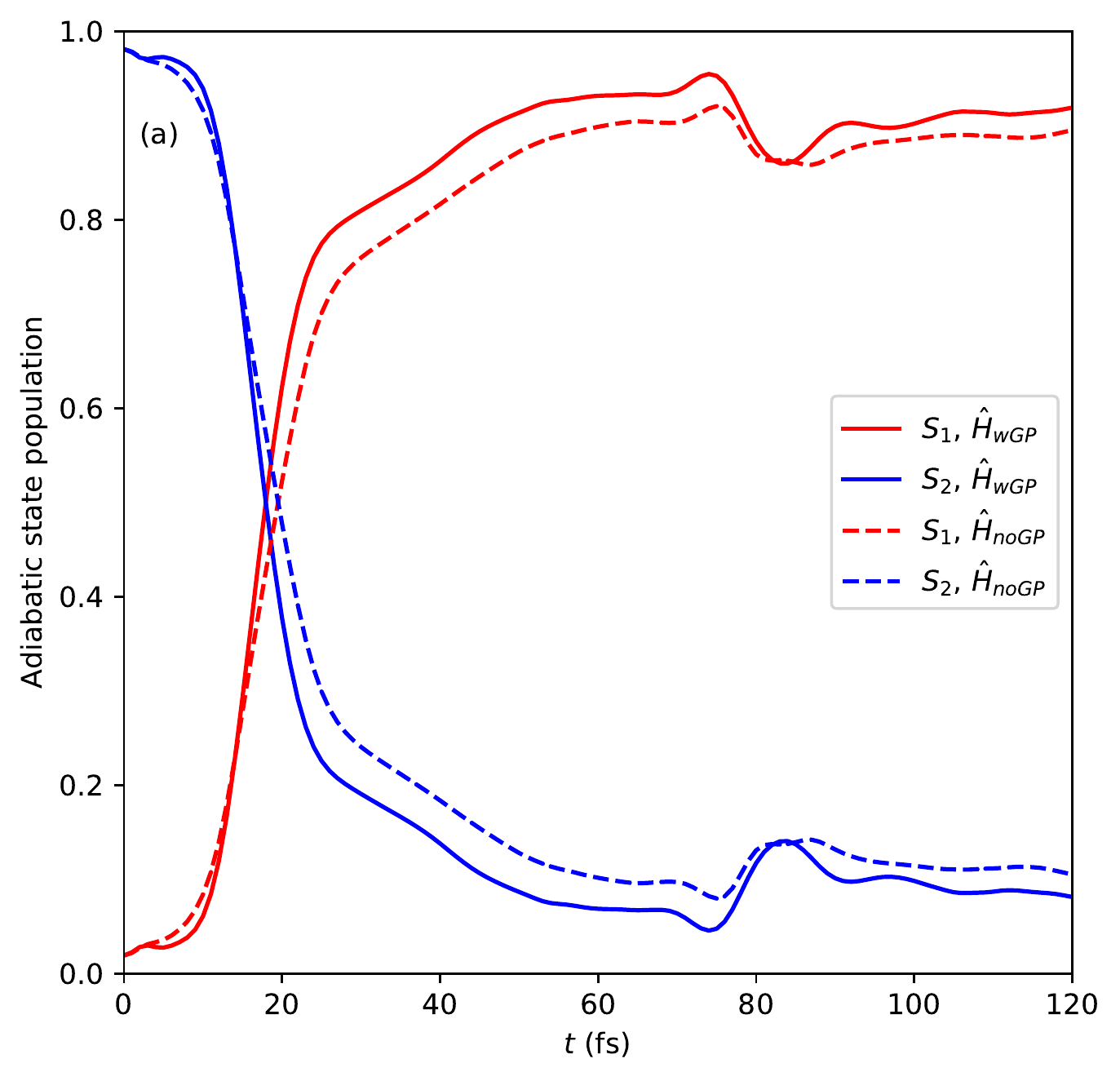}
    \includegraphics[width=7.0cm,angle=0]{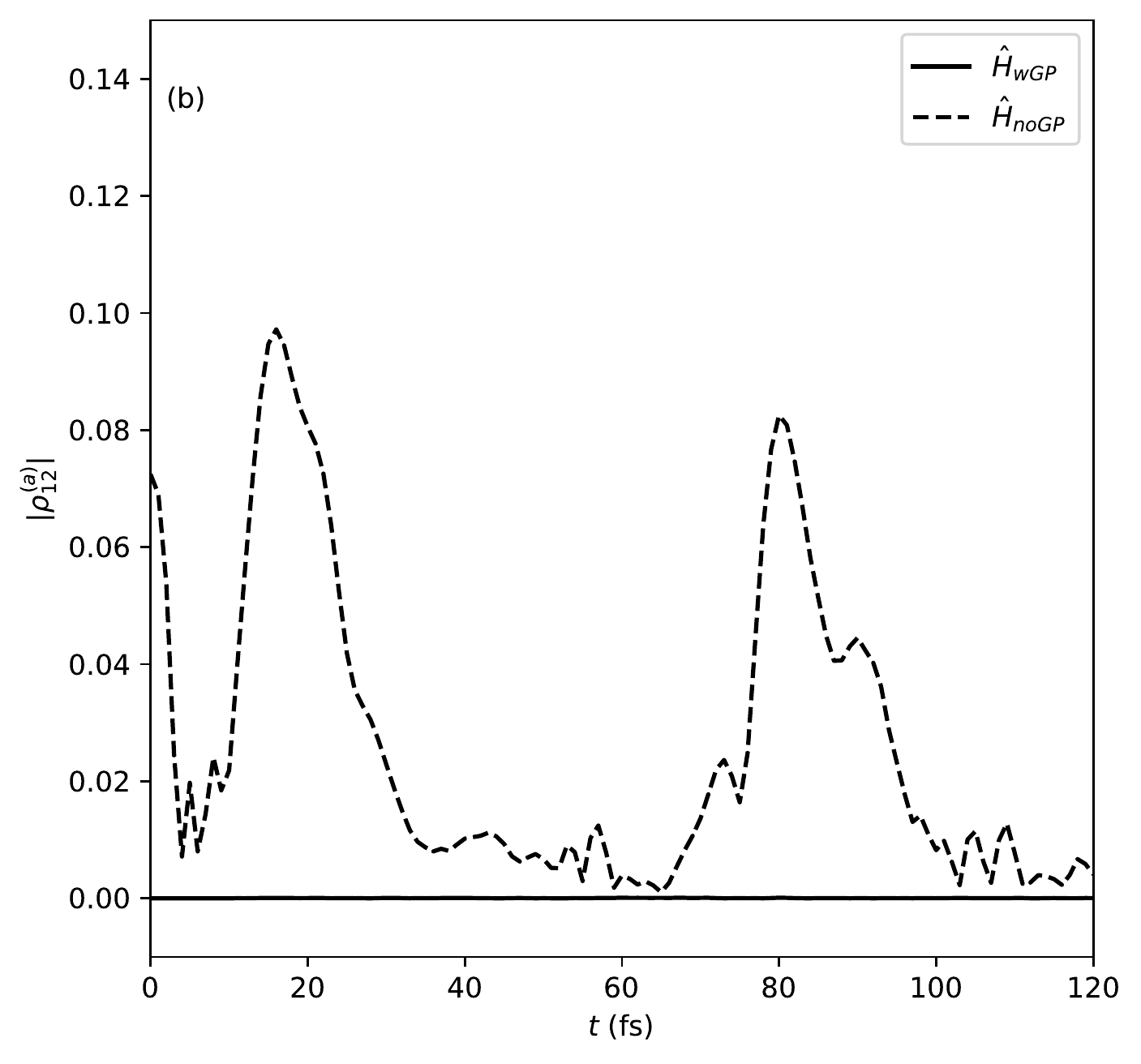}
    \caption{Quantum dynamics of pyrazine following vertical
      excitation to the optically bright $B_{2u}(\pi \pi^{*})$ using
      $\hat{H}_{\text{wGP}}$ (solid lines) and $\hat{H}_{\text{noGP}}$
      (dashed lines). (a) Adiabatic state populations. (b) Electronic
      coherences. It can clearly be seen that omission of the
      geometric phase leads, incorrectly, to large electronic
      coherences which vanish when geometric phase is properly
      included.}
    \label{fig:pyrazine}
  \end{center}
\end{figure*}

%

\end{document}